\documentclass[pra,twocolumn,showpacs,floatfix,a4paper,superscriptaddress]{revtex4}
\usepackage{bm,color,graphicx,amsmath}

\begin{document}

\title{The prospect of detecting single-photon force effects in cavity optomechanics}

\author{H. X. Tang}
\affiliation{Department of Electrical Engineering, Yale University, New Haven, CT 06520, USA}
\author{D.~Vitali}
\affiliation{School of Science and Technology, Physics Division, University of Camerino, via Madonna delle Carceri, 9, I-62032 Camerino (MC), Italy, and INFN, Sezione di Perugia, Italy}

\begin{abstract}
Cavity optomechanical systems are approaching a strong-coupling regime where the coherent dynamics of nanomechanical resonators can be manipulated and controlled by optical fields at the single photon level. Here we propose an interferometric scheme able to detect optomechanical coherent interaction at the single-photon level which is experimentally feasible with state-of-the-art devices.
\end{abstract}


\date{\today}
\maketitle

\section{Introduction}

Ground state cooling of an engineered mechanical resonator has been recently achieved in opto- and electro-mechanical systems exploiting the so called linearized regime where the effective optomechanical interaction is enhanced by strongly driving the selected cavity mode~\cite{TeuflGSC,PainterGSC,KippenbergGSC}. In this regime the system dynamics is linear and one is typically restricted to the manipulation and detection of Gaussian states of optical and mechanical modes~\cite{Hammerer2012}. If the single-photon optomechanical coupling is large enough, the nonlinear dispersive nature of the radiation-pressure interaction would allow the observation of a number of phenomena which has been recently predicted, such as photon blockade~\cite{Rabl2011}, generation of mechanical non-Gaussian steady states~\cite{Nunnenkamp2011,Xu2013}, nontrivial photon statistics in the presence of coherent driving~\cite{Liao2012,Xu2013b,Kronwald2013}, quantum non-demolition measurement~\cite{Ludwig2012} and quantum gates~\cite{Stannigel2012} at the single photon/phonon level. A further interesting direction is to use single photon optomechanical interferometry in this strong coupling regime for generating and detecting quantum superpositions at the macroscopic scale, eventually exploiting post-selection~\cite{Pepper2012,Vanner2013,Akram2013,Hong2013,Sekatski2014,Galland2014}.

The force induced by a photon trapped in a cavity can be described by $f_{\rm ph}=\hbar g_{\rm om}$\cite{PainterZipper}, where an optomechanical coupling strength $g_{\rm om}$ as high as $2\pi \times 123$ GHz/nm has been realized \cite{PainterZipper,Chan2012}. This corresponds to an instantaneous force of $80$ fN applied by one cavity photon. For propagating waves \cite{MoNature08}, the corresponding force per photon is expressed as $F_0 = (\hbar\omega_0/n_{\rm eff})(dn_{\rm eff}/dz)$, where $n_{\rm eff}$ is the effective index of the suspended waveguide, $z$ is the evanescent coupling gap. In a substrate coupled device, $dn_{\rm eff}/dz$ could reach $10^{-3}/$nm~\cite{PerniceOpEx09}. Hence the force per photon could be as high as 130 fN, which is rather significant, since force down to the sub-attonewton level ($10^{-18}N$) has been measured in the literature~\cite{RugarAttoNewton}. Despite its large value, it is experimentally challenging to quantify such an instantaneous force. The interaction time of such a force is the photon cavity lifetime, which is 5 ns in a cavity with mechanical quality factor $Q = 10^6$, and ~7 fs for a propagating photon along a 10 $\mu$m waveguide.  The momentum gained by the mechanical resonator is often weak, on the order of $10^{-22}$ kg $\cdot \mathrm{m/s}$. To measure such a small effect, a compromise has to be made: on one hand, a very compliant resonator is required to improve the responsibility to a weak force; on the other hand, the thermal noise in a compliant resonator can be far more stronger than the photon effect due to the low oscillating frequency. A first theoretical study of the dynamics of an optomechanical cavity in the presence of driving at the single photon level has been carried out in Ref.~\cite{He2012}.

In this paper, we propose an experimental scheme where a very compliant resonating cantilever is coupled to a high Q optical cavity able to achieve sensitive readout of an optomechanical force at the single photon level. The contribution of photon kicks, quantum fluctuations and thermal brownian excitations are calculated based on a Wigner function approach. We found that by performing repeated measurements, the effect of single photons on the cantilever motion can be discriminated from the effect of thermal noise. The desired device parameters along with the requirements on cryogenic cooling are discussed.

\section{The experimental scheme}

The general experimental setup we are considering is formed by an optical cavity which is coupled to a mechanical oscillator, initially at thermal equilibrium at temperature $T$ (see Fig.~1). We consider an experiment in which a first single-photon pulse is sent into the cavity in order to excite the resonator coupled to it. After a variable delay, a second optical pulse driving a different cavity mode is sent into the cavity in order to probe the induced mechanical motion. The proposed experiment is practically an interferometric scheme in which coherence between the two optical pulses is provided by the coherent motion of the oscillator within the two pulses. We will see that if the optomechanical coupling and the mechanical quality factor $Q$ are large enough, the coherent mechanical dynamics can be detected as a modulation of the intensity of the second pulse as a function of the time delay between the two pulses. The oscillating signal is a direct signature of the kick of the single photon pulse and can be viewed as a sort of interference fringes associated with it. Let us describe in detail the three stages of the experiment.

\subsection{First step: the first weak pulse drives the cavity}

The effective Hamiltonian of the system is
\begin{eqnarray}\label{eq:ham1}
   && H_1=\hbar \omega_{C1} a_1^{\dagger}a_1+\hbar \omega_m b^{\dagger}b-\hbar G_1 a_1^{\dagger}a_1(b+b^{\dagger}) \nonumber \\
   &&-i\hbar E_1\left(a_1 e^{-i\omega_{L1}t}-
    a_1^{\dagger} e^{i\omega_{L1}t}\right),
\end{eqnarray}
where $a_1$ is the annihilation operator of the cavity mode driven by the laser pulse, $\omega_{C1}$ denotes its frequency, and $2 \kappa_1$ its bandwidth, while
the nanomechanical resonator has frequency $\omega_m$ and annihilation operator $b$.
$G_1$ is the optomechanical coupling between the driven cavity mode and the resonator, which can be written as
$G_1=F_1/\sqrt{2\hbar m \omega_m}$, where $F_1$ is the force per photon exerted by the cavity field, and $m$ is the mass of the nanomechanical
resonator~\cite{footnote}. The last term in the Hamiltonian describes the driving of the cavity mode by the first laser pulse, with central frequency $\omega_{L1}$ and
with a driving strength described by the rate $E_1$, which is related to the input power $P_1^{in}$ by $E_1=\sqrt{2P_1^{in}\kappa_1/\hbar
\omega_{L1}}$.

A basic assumption for the proposed experiment is considering the bad cavity limit of unresolved sidebands, i.e., $\kappa_1 \gg \omega_m$: this allows to choose a duration $t_1$ of the first pulse such that $\kappa_1 t_1 \gg 1$ and $\omega_m t_1 \ll 1$, which means that during the first pulse the cavity mode reaches its steady state, while the mechanical resonator does not appreciably move. We also assume that the mechanical quality factor $Q_m=\omega_m/\gamma_m$ is sufficiently large ($Q_m \gg 1$), so that also mechanical damping effects are negligible during the first pulse. Therefore the dynamics of the system during the pulse is driven by the Hamiltonian of Eq.~(\ref{eq:ham1}) and by the dissipative term
describing the loss of photons by the cavity due to its nonzero bandwidth. The time evolution of the density matrix $\rho_{1b}$ of the system
is therefore given by the master equation
\begin{equation}\label{eq:meq1}
    \dot{\rho}_{1b}=-\frac{i}{\hbar}\left[H_1,\rho_{1b}\right]+\kappa_1\left(2a_1\rho_{1b}a_1^{\dagger}-a_1^{\dagger}a_1\rho_{1b}-
    \rho_{1b}a_1^{\dagger}a_1\right).
\end{equation}

\begin{figure}[t!]
\begin{center}
\includegraphics[width=0.45\textwidth]{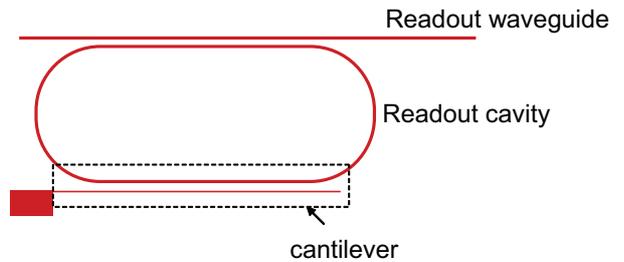}
\caption{Proposed device geometry. The ring cavity, critically coupled to the input/output waveguide, has a quality factor ranging from 10,000 to 100,000. The nanomechanical cantilever is fully suspended and side-coupled to the ring oscillator. The cantilever cross section is designed to be small in order to reduce the spring constant and meanwhile to avoid carrying an optical mode. } \label{fig1}
\end{center}
\end{figure}

\noindent It is convenient to adopt the interaction picture with respect to $H_0=\hbar \omega_{L1}a_1^{\dagger}a_1$, i.e., to move to the frame rotating
at the pulse central frequency $\omega_{L1}$. In this frame, the master equation becomes
\begin{equation}\label{eq:meq2}
    \dot{\rho}_{1b}=-\frac{i}{\hbar}\left[H_1^{ip},\rho_{1b}\right]+\kappa_1\left(2a_1\rho_{1b}a_1^{\dagger}-a_1^{\dagger}a_1\rho_{1b}-
    \rho_{1b}a_1^{\dagger}a_1\right),
\end{equation}
with
\begin{equation}\label{eq:ham2}
    H_1^{ip}=\hbar \Delta_1 a_1^{\dagger}a_1+\hbar \omega_m b^{\dagger}b-\hbar G_1 a_1^{\dagger}a_1 q -i\hbar E_1\left(a_1 - a_1^{\dagger} \right),
\end{equation}
where $\Delta_1=\omega_{C1}-\Omega_{L1}$ is the cavity detuning, and we have defined $q=b+b^{\dagger}$, a dimensionless position operator of the
mechanical resonator.

Due to the assumption $\omega_m t_1 \ll 1$, it is possible to neglect the effect of the free mechanical Hamiltonian $\hbar \omega_m b^{\dagger}b$ in Eq.~(\ref{eq:ham2}) during the first pulse, so that the dynamics become easy to solve. In this limit, in fact, the resonator position $q$ is a constant of motion, and it is therefore convenient to work in the position eigenstate basis, $|q\rangle$. To be more specific, assuming a factorized initial state $\rho_{1b}(0)=\rho_b(0)\otimes \rho_1(0)$, and writing the initial state of the
mechanical resonator in the position eigenstate basis as $\rho_b(0)=\int dq dq' \rho_b(0,q,q')|q\rangle \langle q'|$, the state of the whole system at the end of the first pulse can be written as
\begin{equation}\label{eq:state1}
    \rho_{1b}(t_1)=\int dq dq' \rho_b(0,q,q')|q\rangle \langle q' |\otimes K_1(q,q',t_1)\left[\rho_1(0)\right],
\end{equation}
where $K_1(q,q',t_1)$ is the time evolution superoperator acting only on the cavity mode and where $q$ and $q'$ are real valued parameters. This
time evolution is simple and describes a driven cavity mode, with decay rate $\kappa_1$ and detuning
$\Delta_1(q)=\Delta_1-G_1 q$. In the present case, the initial state is given by a thermal
equilibrium state at temperature $T$, $\rho_b(0)=\rho_b^{th}$ for the mechanical resonator, and by the vacuum state $\rho_1(0)=|0\rangle \langle 0|$
for the cavity mode. The dynamics determined by the superoperator $K_1(q,q',t_1)$ preserves coherent states, that is, transforms an initial
coherent state into another coherent state, and it is easy to verify that
\begin{equation}\label{eq:state1b}
    K_1(q,q',t_1)\left[|0\rangle \langle 0|\right]=|\alpha_1(q,t_1)\rangle \langle \alpha_1(q',t_1)|,
\end{equation}
where $|\alpha_1(q,t_1)\rangle $ is the coherent state of mode $a_1$ with amplitude
$$
\alpha_1(q,t_1)=\frac{E_1}{\kappa_1+i(\Delta_1-G_1 q)}\left[1-e^{-(\kappa_1+i\Delta_1-iG_1 q)t_1}\right].
$$
Since we have assumed $\kappa_1 t_1 \gg 1$, one can well approximate the amplitude at the end of the first pulse with
its asymptotic value,
\begin{equation}\label{eq:asy}
\alpha_1(q,t_1)\simeq \alpha_1(q)=\frac{E_1}{\kappa_1+i(\Delta_1-G_1 q)}.
\end{equation}
Therefore the state of the optomechanical system at the end of the first pulse can be written as
\begin{equation}\label{eq:state2}
    \rho_{1b}(t_1)=\int dq dq' \rho_b^{th}(q,q')|q\rangle \langle q' |\otimes |\alpha_1(q)\rangle \langle \alpha_1(q')|,
\end{equation}
where
\begin{equation}\label{eq:state2b}
    \rho_b^{th}(q,q')= \frac{\exp\left[-\frac{(q+q')^2}{8(1+2 \bar{n})}-\frac{(q-q')^2(1+2
    \bar{n})}{8}\right]}{\sqrt{2\pi (1+2 \bar{n})}}
\end{equation}
is the coordinate representation of the thermal equilibrium state of the mechanical oscillator \cite{Landau}, with mean thermal vibrational
number $\bar{n}=\left[\exp\left(\hbar \omega_m/k_B T\right)-1\right]^{-1}$. Since we want to test the sensitivity of the optomechanical device at single-photon level, we assume a very weak resonant driving pulse, that is we choose $\Delta_1 =0$, and a mean number of cavity photons at the end of the pulse, $|\alpha_1(q,t_1)|^2 \simeq E_1^2/\kappa_1^2 \simeq 1$.

After the first pulse, cavity mode $a_1$ quickly decays to the vacuum state in a time roughly equal to
$1/\kappa_1$, with a negligible effect on the motion of the mechanical resonator, whose reduced state at the end of the first pulse is obtained by tracing out mode $a_1$ in Eq.~(\ref{eq:state2}),
\begin{eqnarray}\label{eq:state3}
  \rho_{b}^{red}(t_1)&=& \int dq dq' \rho_b^{th}(q,q')\langle \alpha_1(q')|\alpha_1(q)\rangle |q\rangle \langle q'| \nonumber \\
   & \equiv &
    \int dq dq' \rho_1(q,q') |q\rangle \langle q'|,
\end{eqnarray}
where
$$
\langle \alpha_1(q')|\alpha_1(q)\rangle = \exp\left[-\frac{|\alpha_1(q)|^2+|\alpha_1(q')|^2}{2}+\alpha_1^*(q') \alpha_1(q)\right]
$$
is the overlap between the two coherent states. Using Eqs.~(\ref{eq:asy}) and (\ref{eq:state2b}) and exploiting the fact that $\Delta_1=0$, one has that the state of the mechanical resonator at the end of the first pulse is equal to
\begin{widetext}
\begin{equation}\label{eq:rho1}
    \rho_1(q,q')=\frac{1}{\sqrt{2\pi (1+2 \bar{n})}}\exp\left[-\frac{(q+q')^2}{8(1+2 \bar{n})}-\frac{(q-q')^2}{8}(1+2
    \bar{n})\right] \exp\left\{-\frac{|E_1|^2\left[G_1^2(q-q')^2-2iG_1
    \kappa_1(q-q')\right]}{2\left[\kappa_1^2+G_1^2q^2\right]\left[\kappa_1^2+G_1^2q'^2\right]}\right\}.
\end{equation}
\end{widetext}
Due to the assumptions above, the optomechanical interaction during the first pulse does not change the probability distribution of the oscillator position
$q$. This interaction however changes the \emph{momentum distribution} of the
oscillator, as manifested by the modification of the off-diagonal terms ($q \neq q'$) of the density matrix of the mechanical oscillator. This
modification is more evident if we represent such a state in the phase space picture provided by the Wigner function, which is related to the position representation of the density matrix by the following Fourier transform \cite{Gardiner}
\begin{equation}\label{eq:Wig-def}
    W(q,p)=\frac{1}{2\pi}\int dy \rho(q+y,q-y) e^{-i y p}.
\end{equation}
If we insert Eq.~(\ref{eq:rho1}) into this definition we get a non Gaussian integral, which will be approximately evaluated in the following section.

\subsection{Second step: free dynamics between the two pulses}

The adoption of the Wigner function is useful also because the dynamics during the second step of the experiment is
easier to describe in this picture. During this step, which lasts a time interval $\tau$, the cavity remains empty because mode $a_1$ is back in its vacuum state and no other mode is excited. Therefore, only the mechanical resonator evolves, driven solely by its free Hamiltonian $H_{m}=\hbar
\omega_m b^{\dagger} b$, because we have assumed $Q_m \gg 1$, implying that dissipative effects are negligible as long as $\gamma_m \tau \ll 1$.
The corresponding time evolution in the Wigner representation is simply a rotation in phase space by an angle $\omega_m
\tau$. In fact, denoting with $W_1(q,p)$ the Wigner function associated with the state of the resonator at the end of the first pulse of Eq.~(\ref{eq:rho1}), and with $W_f(q,p)$ the Wigner function at time $t_1+\tau$, just before the second pulse, one has
\begin{eqnarray}\label{eq:wig-evol}
  &&  W_f(q,p)= W_1(q_{\tau},p_{\tau}) \\
   && \equiv W_1(q\cos \omega_m \tau-p\sin \omega_m \tau,p\cos \omega_m \tau+q\sin \omega_m \tau). \nonumber
\end{eqnarray}
The state of the resonator just before the second pulse can be equivalently written in the position representation by taking the inverse Fourier transform of Eq.~(\ref{eq:Wig-def}), getting
\begin{equation}\label{eq:Wig-antidef}
    \rho_f(q+y,q-y)=\int dp W_1(q_{\tau},p_{\tau})e^{i y p}.
\end{equation}

\subsection{Third step: the second probe pulse is sent into the cavity and its intensity is detected}

The second pulse is quasi-resonant with a second, different mode, of the cavity, with annihilation operator $a_2$. However we assume again the bad cavity limit of unresolved sidebands and we again choose a pulse duration $t_2$ such that $\kappa_2 t_2 \gg 1$ and $\omega_m t_2 \ll 1$. Therefore the dynamics is
identical to that taking place during the first pulse, with index $1$ replaced by index $2$, associated with the second probe mode. In
the interaction picture with respect to $H_0=\hbar \omega_{L2}a_2^{\dagger}a_2$, i.e., in the frame rotating at the pulse central frequency
$\omega_{L2}$ the dynamics of the density operator of the system formed by cavity mode 2 and by the mechanical resonator is now generated by the
master equation
\begin{equation}\label{eq:meq22}
    \dot{\rho}_{2b}=-\frac{i}{\hbar}\left[H_2^{ip},\rho_{2b}\right]+\kappa_2\left(2a_2\rho_{2b}a_2^{\dagger}-a_2^{\dagger}a_2\rho_{2b}-
    \rho_{2b}a_2^{\dagger}a_2\right),
\end{equation}
where
\begin{equation}\label{eq:ham22}
    H_2^{ip}=\hbar \Delta_2 a_2^{\dagger}a_2+\hbar \omega_m b^{\dagger}b-\hbar G_2 a_2^{\dagger}a_2 q -i\hbar E_2\left(a_2 - a_2^{\dagger} \right),
\end{equation}
where $\Delta_2=\omega_{C2}-\Omega_{L2}$ is the detuning of the second cavity mode. The parameters associated with this second mode, $\Delta_2$,
$\kappa_2$, $G_2$, $E_2$ and the duration of the second pulse $t_2$ are generally different from those associated with the first mode, but their
order of magnitude is just the same (apart from $E_2$ which is now much larger because the probe pulse is more intense than the first). In
particular, since the initial state of mode 2 is again the vacuum state $|0 \rangle \langle 0|$, one can repeat exactly the same
calculation and approximations made in step 1. This means that the oscillator does not move during the second pulse and
that also mode $2$ is always in a coherent state. Therefore, the state of the system formed by the cavity mode 2 and the mechanical beam, at the end of the second pulse, can be written as
\begin{equation}\label{eq:state12}
    \rho_{2b}(t_1+\tau+t_2)=\int dq dq' \rho_f(q,q')|q\rangle \langle q' |\otimes |\alpha_2(q)\rangle \langle \alpha_2(q')|,
\end{equation}
where $\rho_f(q,q')$ is just the state in the position representation of Eq.~(\ref{eq:Wig-antidef}), and $|\alpha_2(q)\rangle $ is the coherent state of mode $a_2$ with amplitude
$$
\alpha_2(q)=\frac{E_2}{\kappa_2+i(\Delta_2-G_2 q)}.
$$
The measured observable is the number of photons in mode $2$, $\langle a_2^{\dagger}a_2\rangle $ which, using the final state of
Eq.~(\ref{eq:state12}) and Eq.~(\ref{eq:Wig-antidef}), can be written as
\begin{eqnarray}\label{eq:nphot2}
  \langle a_2^{\dagger}a_2\rangle &=& {\rm Tr}\left\{a_2^{\dagger}a_2  \rho_{2b}(t_1+\tau+t_2)\right\} \nonumber \\
  &=&  \int dq  \rho_f(q,q)|\alpha_2(q)|^2 \\
  &=& \int \int dq dp W_1(q_{\tau},p_{\tau})\frac{|E_2|^2}{\kappa_2^2+(\Delta_2-G_2 q)^2}. \nonumber
\end{eqnarray}
This expression has an intuitive explanation since it shows that the intensity of the second pulse is given by the usual Lorenzian response of the cavity, with a detuning modulated by the resonator motion, averaged over the pertinent phase-space distribution of the mechanical resonator. Eq.~(\ref{eq:nphot2}) also illustrates the basic idea of the proposed experiment: in fact, the detection of oscillations of $\langle a_2^{\dagger}a_2\rangle$ as a function of $\omega_m \tau$ with period equal to $2 \pi$ would be an evident signature of the effect of the force imparted by the single-photon pulse. In fact, the resonator position will oscillate at angular frequency $\omega_m$, consequently modulating the detuning and the intracavity photon number. The average over the phase-space distribution of the mechanical resonator implies that the ``fringe'' visibility of these oscillations may be blurred by thermal effects. We expect in fact that the time modulation of the signal (i.e., of the intracavity photon number) is visible only if the momentum kick provided by the first single-photon pulse emerges from the average over the thermal distribution, and this is possible only if the single-photon optomechanical coupling $G_1$ is large enough.

\begin{figure}[htb]
\begin{center}
\includegraphics[width=.49\textwidth]{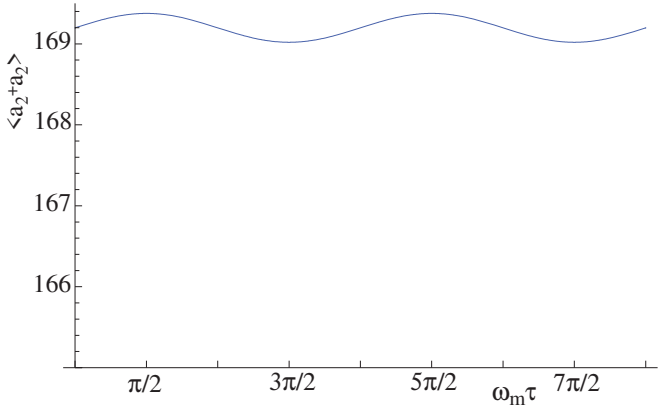}
\caption{Mean number of photons in the second probe pulse versus $\omega_m \tau$. Parameter values are $\kappa_1=\kappa_2=1$ GHz, $E_1/\kappa_1 = 1$ (i.e.,
mean number of photons in the cavity at the end of the first pulse equal to one), $m=1$ pg, $\omega_m/2\pi = 10$ MHz, $F_1=F_2=0.1$ pN, implying single photon couplings $G_1=G_2=0.03$ GHz; the mean thermal number of mechanical quanta is $\bar{n}=10^4$, corresponding to $T\simeq 6$ K, while for the second pulse we have chosen $\Delta_2=20$ GHz and $E_2=1$ THz which, for pulses of duration $t_1 \simeq t_2 \simeq 5$ ns, means sending about $2.5 \times 10^6$ photons in the second pulse. The fringe visibility in this case is approximately equal to $10^{-3}$.} \label{fig2}
\end{center}
\end{figure}

Eq.~(\ref{eq:nphot2}) can be explicitly calculated in terms of the system parameters, by first changing integration variables $(q,p) \to (q_{\tau},p_{\tau})$ and explicitly writing $W_1(q_{\tau},p_{\tau})$ in terms of the position representation of the resonator state after the first pulse $\rho_1(q,q')$ of Eq.~(\ref{eq:rho1}). Using $dp dq = dq_{\tau}dp_{\tau}$ and $q=q_{\tau}\cos \omega_m\tau +p_{\tau}\sin \omega_m \tau$, one finally gets
\begin{widetext}
\begin{equation}\label{eq:finalgen}
 \langle a_2^{\dagger}a_2\rangle \equiv \bar{n}_2(\omega_m\tau)=\int \int \int dq_{\tau} dp_{\tau} dy \rho_1(q_{\tau}+y,q_{\tau}-y)e^{-i y p_{\tau}}\frac{|E_2|^2}{\kappa_2^2+(\Delta_2-G_2 q_{\tau}\cos \omega_m\tau -G_2 p_{\tau}\sin \omega_m \tau)^2},
\end{equation}
where $\rho_1(q_{\tau}+y,q_{\tau}-y)$ is explicitly given by by Eq.~(\ref{eq:rho1}),
\begin{equation}\label{eq:rho1qy}
    \rho_1(q_{\tau}+y,q_{\tau}-y)=\frac{1}{\sqrt{2\pi (1+2 \bar{n})}}\exp\left[-\frac{q_{\tau}^2}{2(1+2 \bar{n})}-\frac{y^2}{2}(1+2
    \bar{n})\right] \exp\left\{-\frac{2|E_1|^2\left[G_1^2 y^2-iG_1 \kappa_1 y\right]}{\left[\kappa_1^2+G_1^2(q_{\tau}+y)^2\right]\left[\kappa_1^2+G_1^2(q_{\tau}-y)^2\right]}\right\}.
\end{equation}
\end{widetext}
Eqs.~(\ref{eq:finalgen})-(\ref{eq:rho1qy}) can be evaluated numerically and provide the desired signal $\bar{n}_2(\omega_m\tau)$ under the only assumptions $\kappa_j t_j \gg 1$ and $\omega_m t_j \ll 1$, $ j=1,2$. An explicit example is shown in Fig.~\ref{fig2}, which corresponds to the parameter set illustrated in the figure caption. The time modulation of the intensity of the second pulse is rather weak, with a fringe visibility of the order of $10^{-3}$, which however should be visible by acquiring sufficiently large statistics.

\section{Analysis of the results}
The dependence of the visibility upon the various system parameters, and the conditions under which the signal modulation due to the resonator motion caused by the first single-photon kick can be better understood by approximately evaluating the general expressions of Eqs.~(\ref{eq:finalgen})-(\ref{eq:rho1qy}). In fact, one can derive a much simpler expression for $\langle a_2^{\dagger}a_2\rangle$ in the limit of not too strong single-photon coupling, $G_1/\kappa_1 \ll 1$. This condition is quite easily satisfied and in fact, the parameters of Fig.~2 correspond to $G_1/\kappa_1 =0.03$. This assumption allows to simplify the expression of the exponent in Eq.~(\ref{eq:rho1}). In fact, at second
order in $G_1/\kappa_1$ we get
\begin{widetext}
\begin{equation}\label{eq:rho1bis}
    \rho_1(q+y,q-y)\simeq \frac{1}{\sqrt{2\pi (1+2 \bar{n})}}\exp\left[-\frac{q^2}{2(1+2 \bar{n})}-\frac{y^2}{2}(1+2
    \bar{n})\right] \exp\left\{-\frac{2|E_1|^2}{\kappa_1^2}\left[\frac{G_1^2 y^2}{\kappa_1^2}-i\frac{G_1 y}{\kappa_1}\right]
    \right\}.
\end{equation}
\end{widetext}
As a consequence, in this limit, using the definition of Eq.~(\ref{eq:Wig-def}), the Wigner function of the resonator at the end of the first pulse can be easily evaluated, and one gets
\begin{equation}\label{eq:rho1wig}
    W_1(q,p)= \frac{1}{\pi\sqrt{2(1+2 \bar{n})a_1}}\exp\left[-\frac{q^2}{2(1+2 \bar{n})}-\frac{\left(p-p_0\right)^2}{a_1}\right],
\end{equation}
where
\begin{eqnarray}
  p_0 &=& \frac{2|E_1|^2 G_1}{\kappa_1^3}, \\
  a_1 &=& 2(1+2 \bar{n})+\frac{8|E_1|^2 G_1^2}{\kappa_1^4}. \label{eq:a1}
\end{eqnarray}
Eq.~(\ref{eq:rho1wig}) shows that, due to the kick exerted by the first pulse, the mechanical beam acquires a nonzero mean momentum
$p_0$. At second order in $G_1^2/\kappa_1^2$ one has also an increase of the width of the momentum probability distribution with respect to the
thermal width (see Eq.~(\ref{eq:a1})). However this correction is negligible with respect to the thermal width even at zero temperature $\bar{n}=0$, because in the experimental regime we are considering $|E_1|^2/\kappa_1^2 \sim 1$ and $ G_1^2/\kappa_1^2 \ll 1$.
Therefore a first order treatment in $G_1/\kappa_1$ is well justified and we shall assume $a_1 \simeq 2(1+2
\bar{n})$ from now on. This implies that at the end of the first pulse the state of the mechanical resonator is, with a good approximation,
equal to a thermal state displaced along the momentum axis by the nonzero mean momentum $p_0$.

Then, using the fact that between the two pulses one has the free oscillation of the mechanical resonator which is a rotation by an angle $\omega_m\tau$ in phase space, and taking the inverse Fourier transform,
one gets for the position representation of the state of the mechanical resonator before the second pulse
\begin{widetext}
\begin{equation}\label{eq:rhof}
    \rho_f(q+y,q-y)= \frac{1}{\sqrt{2\pi (1+2 \bar{n})}}\exp\left[-\frac{(q-p_0 \sin\omega_m \tau)^2}{2(1+2 \bar{n})}-\frac{y^2}{2}(1+2
    \bar{n})+i y p_0 \cos \omega_m \tau \right].
\end{equation}
\end{widetext}
The probability distribution of the resonator position just before the second pulse is immediately obtained putting $y=0$ in Eq.~(\ref{eq:rhof}): as expected, the momentum kick received by the mechanical resonator during the first pulse manifests itself as an oscillation with frequency $\omega_m$ of the center of the
probability distribution. However, due to the condition $\omega_m t_2 \ll 1$ this probability distribution of the resonator position is still valid at the end of the second pulse, and therefore inserting it in the general equation for the detected signal of Eq.~(\ref{eq:nphot2}), one gets
\begin{widetext}
\begin{equation}\label{eq:nphot3}
  \langle a_2^{\dagger}a_2\rangle = \frac{1}{\sqrt{2\pi (1+2 \bar{n})}}\int dq  \exp\left[-\frac{(q-p_0 \sin\omega_m \tau)^2}{2(1+2 \bar{n})}\right]
  \frac{|E_2|^2}{\kappa_2^2+(\Delta_2-G_2 q)^2},
\end{equation}
which is much simpler than the general expression of Eq.~(\ref{eq:finalgen}) obtained without assuming also $G_1/\kappa_1 \ll 1$. Changing integration variable we can also rewrite it as
\begin{equation}\label{eq:fin-gen-app4}
  \langle a_2^{\dagger}a_2\rangle =\int
   \frac{dq}{\sqrt{2\pi (1+2 \bar{n})}}
   \exp\left\{-\frac{ q^2}{2 (1+2 \bar{n})}\right\}
  \frac{|E_2|^2}{\kappa_2^2+\left[\Delta_2(\tau)-G_2 q\right]^2},
\end{equation}
\end{widetext}
where we have introduced an effective time-dependent detuning
\begin{equation}\label{eq:det-eff}
\Delta_2(\tau)=\Delta_2-G_2 p_0 \sin \omega_m \tau=\Delta_2-\frac{2G_1 G_2 |E_1|^2}{\kappa_1^3}\sin \omega_m \tau.
\end{equation}
This simpler expression for the intensity of the probe pulse is the convolution of a Gaussian probability distribution (which comes from the
thermal distribution of the oscillator position) with the Lorenzian curve associated with the cavity mode response. This convolution is the
well-known Voigt profile function in spectroscopy.

From this expression one can estimate how the fringe visibility is influenced by the various system parameters. The Voigt function is well approximated by a
Lorenzian or by a Gaussian depending upon the ratio between the width of the two curves. If the Gaussian width (given by the thermal
equilibrium width $\sqrt{1+2 \bar{n}}$) is much larger than the Lorenzian width (given by $\kappa_2/G_2$) then Eq.~(\ref{eq:fin-gen-app4}) is
well approximated by a Gaussian, because one can replace the Lorenzian with a suitably normalized Dirac delta function. In the opposite case
the role of the Gaussian and Lorenzian curves are exchanged. When $1+2 \bar{n} \gg \kappa_2^2/G_2^2$, the Gaussian distribution is wider and we can approximate
\begin{equation}\label{eq:fin-gen-app5}
  \langle a_2^{\dagger}a_2\rangle \simeq \frac{|E_2|^2}{\kappa_2 G_2 }
   \frac{\sqrt{\pi}}{\sqrt{2 (1+2 \bar{n})}}
   \exp\left\{-\frac{ \Delta_2(\tau)^2}{2G_2^2 (1+2 \bar{n})}\right\}.
\end{equation}
This equation reproduces the correct ``classical'' limit, which we mean as the limit of very large temperatures and at the same time very large
intensity of the second driving pulse, i.e., $\bar{n} \to \infty$ and $|E_2| \to \infty$, but with $|E_2|^2/\sqrt{\bar{n}} \to $ constant. In
such a limit the detected probe intensity becomes a constant and does not oscillate, that is, the fringe visibility is equal to zero. An explicit expression of the fringe visibility $V$ can be obtained in this limit, using Eq.~(\ref{eq:det-eff}) for $\Delta_2(\tau)$: one gets
\begin{eqnarray} \label{eq:visib}
    && V \equiv \frac{ \langle a_2^{\dagger}a_2\rangle_{max}-\langle a_2^{\dagger}a_2\rangle_{min}}
    {\langle a_2^{\dagger}a_2\rangle_{max}+\langle a_2^{\dagger}a_2\rangle_{min}}  \\
    &&= \frac{\exp\left\{-\frac{ (\Delta_2-G_2p_0)^2}{2G_2^2 (1+2 \bar{n})}\right\}-\exp\left\{-\frac{ (\Delta_2+G_2p_0)^2}{2G_2^2 (1+2 \bar{n})}\right\}}
    {\exp\left\{-\frac{ (\Delta_2-G_2 p_0)^2}{2G_2^2 (1+2 \bar{n})}\right\}+\exp\left\{-\frac{ (\Delta_2+G_2 p_0)^2}{2G_2^2 (1+2
    \bar{n})}\right\}}\nonumber \\
    &&=\tanh\left[\frac{\Delta_2 p_0}{G_2 (1+2 \bar{n})}\right]=\tanh\left[\frac{2\Delta_2}{\kappa_1}
    \frac{|E_1|^2}{\kappa_1^2}\frac{G_1}{G_2}\frac{1}{1+2\bar{n}}\right]. \nonumber
\end{eqnarray}
Fig.~2 refers to a set of parameters in which we can approximately apply the present Gaussian approximation of the Voigt function. In fact, we have chosen in Fig.~2 $\bar{n}=10^4$ and $\kappa_2^2/G_2^2=\kappa_1^2/G_1^2\simeq 10^3$, and in fact Eq.~(\ref{eq:visib}) gives $V\simeq 10^{-3}$ as found in Fig.~2.

It is interesting to study the opposite limit $1+2 \bar{n} \ll \kappa_2^2/G_2^2$ in which the Gaussian thermal distribution is much narrower than the Lorenzian and one can approximate the former with a Dirac delta in Eq.~(\ref{eq:fin-gen-app4}). Using also the expression of $\Delta_2(\tau)$ of Eq.~(\ref{eq:det-eff}), we get
\begin{equation}\label{eq:fin-gen-app7}
  \langle a_2^{\dagger}a_2\rangle \simeq
  \frac{|E_2|^2}{\kappa_2^2+\left[\Delta_2-G_2 p_0\sin \omega_m \tau\right]^2},
\end{equation}
that is, the intensity of the probe pulse is a Lorenzian modulated in time by the resonator motion caused by the kick imparted by the first pulse. In such a case the visibility is given by
\begin{equation}\label{eq:visib2}
    V=\frac{2 |\Delta_2|G_2 p_0}{\kappa_2^2+ G_2^2p_0^2+
    \Delta_2^2}.
\end{equation}
This visibility is maximized for the following choice of the detuning,
$$|\Delta_2|=\sqrt{\kappa_2^2+ G_2^2p_0^2}$$
which corresponds to the maximum fringe visibility
\begin{equation}\label{eq:visib3}
    V^{max}=\frac{ G_2 p_0}{\sqrt{\kappa_2^2+ G_2^2p_0^2}} .
\end{equation}

We note that the visibility predicted here, although weak, could be measured with the state-of-the-art single photon detectors. To obtain good statistics, 1 million repeated measurements need to be carried out to distinguish a visibility of  $10^{-3}$. At $1$ kHz repetition rate, this corresponds to 20 minutes counting time. The dark count of the detector should be much less than 1000 during this measurement time. Recently such detector has emerged and made available in an integrated photonic circuit~\cite{PerniceDetector}. In fact, quantum efficiency up to 50\% can be reached with dark count rate less than 0.1 Hz, or 100 counts in 20 minutes~\cite{SchuckDetector}. It is foreseeable that a fully integrated device system with on-chip single photon detectors would allow unambiguous determination of single photon effect.

\section{Conclusions}

We have proposed an interferometric scheme employing an experimentally achievable optomechanical device with sufficiently large single photon optomechanical coupling $G/\kappa \sim 10^{-2}$, able to detect and measure the effect of an almost instantaneous optomechanical force exerted by a single photon pulse on a nanomechanical resonator. If the single-photon optomechanical coupling and the mechanical quality factor are sufficiently large, the coherent oscillations of the resonator caused by the kick imparted by a single photon pulse could be detected as time-dependent modulations of the light intensity at the cavity output. This represent a viable scheme able to unambiguously prove optomechanical effects at the single photon level, and which could be efficiently employed also in different scenarios for the interferometric detection of weak impulsive forces in nanomechanical devices.

\section{Acknowledgments}
This work has been supported by the European Commission (ITN-Marie Curie project cQOM and FET-Open Project iQUOEMS), by MIUR (PRIN 2011), and DARPA MTO's ORCHID program through a grant from AFOSR.


\begin{thebibliography}{99}
\bibitem{footnote}The resonator actually interacts also with all the other cavity modes with a similar interaction term. However, these modes are undriven and therefore always remain in their vacuum state. As a consequence, they have no effect on the resonator motion and can be neglected in Eq.~(\ref{eq:ham1}).
\end{thebibliography}
\end{document}